\documentclass[onecolumn]{aastex63}

\newcommand{\smz}{\langle B_z\rangle}
\newcommand{\umz}{\langle \vert B_z\vert\rangle_{\tau=1}}
\newcommand{\avr}[1]{\langle{#1}\rangle}

\usepackage{hyperref}
\usepackage[all]{hypcap}

\begin{document}
\title{On the contribution of quiet Sun magnetism to solar irradiance variations: Constraints on quiet Sun variability and grand minimum scenarios}
\shorttitle{}

\author{M. Rempel}
\affiliation{High Altitude Observatory, NCAR, P.O. Box 3000, Boulder, Colorado 80307, USA}

\shortauthors{Rempel}

\email{rempel@ucar.edu}

\begin{abstract}
While the quiet Sun magnetic field shows only little variation with the solar cycle, long-term variations cannot be completely ruled out from first principles.
We investigate the potential effect of quiet Sun magnetism on spectral solar irradiance through a series of small-scale dynamo simulations with zero vertical 
flux imbalance ($\smz=0$) and varying levels of small-scale magnetic field strength, and one weak network case with an additional flux imbalance 
corresponding to a flux density of $\smz=100$~G. From these setups we compute the dependence
of the outgoing radiative energy flux on the mean vertical magnetic field strength in the photosphere at continuum optical depth $\tau=1$ ($\umz$). We find that a quiet Sun
setup with a mean vertical field strength of $\umz=69$~G is about $0.6~\%$ brighter than a non-magnetic reference case. We find a linear dependence
of the outgoing radiative energy flux on the mean field strength $\umz$ with a relative slope of $1.4\cdot 10^{-4}$~G$^{-1}$. With this sensitivity, only a moderate 
change of the quiet Sun field strength by $10\%$ would lead to a total solar irradiance variation comparable to the observed solar cycle variation. While 
this does provide strong indirect constraints on possible quiet Sun variations during a regular solar cycle, it also emphasizes that potential variability over 
longer time scales could make a significant contribution to longer-term solar irradiance variations.
\end{abstract}


\received{}
\accepted{}

%\maketitle
\section{Introduction}
Reconstructions of the solar total (TSI) and spectral (SSI) irradiance are a crucial input for climate simulations covering time scales of decades to centuries
\citep[see, e.g.][]{Gray:etal:2010}. Of particular interest are reconstructions of the solar irradiance during grand minimum epochs, such as the Maunder minimum during the years
of  1645  to 1715.  Models based on proxy reconstructions and semi-empirical physics have given a wide range of possible irradiance scenarios for the
Maunder minimum ranging  from being comparable to a current day quiet Sun to TSI reductions of as much as 0.4\% (6 W$\,$m$^{-2}$). \citet{Schrijver:etal:2011} 
argued that the Maunder minimum would not differ strongly from a current deep cycle minimum and suggested to use observations from 2008/2009 to constrain 
a possible grand minimum scenario directly. A similar result was also found in a recent reconstruction by \citet{Wu:etal:2018:satire}, who presented a Maunder
minimum TSI Ievel of only 0.2 W$\,$m$^{-2}$ lower than the 2008/2009 cycle minimum. Support for this scenario comes also from the work of \cite{Judge:Saar:2007}, who
studied UV and X-ray emission from flat activity solar-like stars and found emission levels similar to current solar minimum values. Assuming that flat activity stars do 
present a solar-like star in a grand minimum phase, they concluded that a Maunder minimum Sun was likely very similar to a current cycle minimum.
In contrast to this, \citet{Shapiro:etal:2011} and \citet{Egorova:etal:2018} find 
in their reconstructions significantly larger TSI reductions of up to 6 W$\,$m$^{-2}$. \citet{Shapiro:etal:2011} suggested an approach in which a transition from the current
day quiet Sun irradiance to an absolute minimum state of the Sun is based on a proxy derived from cosmogenic isotopes \citep[e.g.][]{Steinhilber:etal:2008}. 
Subsequently it was pointed out by \citet{Judge:etal:2012} that the chosen model for absolute minimum state in \citet{Shapiro:etal:2011} overestimates 
potential irradiance changes by about a factor of two. Furthermore it was pointed out that modulating the quiet Sun contribution with the solar modulation potential
(a measure of the global rather than the local field) is potentially flawed if the quiet Sun magnetic field is of independent origin in terms of a small-scale dynamo
\citep[e.g.][]{Voegler:Schuessler:2007,Rempel:2014:SSD,Kitiashvili:2015:SSD,Khomenko:2017:Biermann}. While \citet{Egorova:etal:2018} took into consideration 
the corrections suggested by \citet{Judge:etal:2012}, they still found a TSI reduction for the Maunder minimum compared to current day quiet Sun in the 3-6 W$\,$m$^{-2}$ 
range, depending on the particular modulation potential considered.  

In this publication we approach the problem from a different angle. We use numerical quiet Sun models in which  mixed polarity small-scale magnetic field is
maintained by a small-scale dynamo in combination with deep recirculation that attempts to capture flux transport from deeper magnetized layers of the convection 
zone that are not directly modeled \citep{Rempel:2014:SSD,Rempel:2018:expl}. Through a variation of the deep recirculation component we can vary the level of quiet Sun 
magnetic field and compute directly the
total and spectral solar irradiance response using an approach similar to \citet{Norris:etal:2017}. While \citet{Norris:etal:2017} considered primarily models with non-zero
net magnetic flux more representative of solar network regions, we focus primarily on setups with zero net magnetic flux that better represent quiet Sun conditions. 
\cite{Yeo:etal:2017:irrad} used the radiative MHD simulations analyzed in \citet{Norris:etal:2017} together with HMI magnetograms to construct an irradiance model that is 
completely independent from the observed TSI record. They were able to capture 95\% of the observed TSI variability in the 2010 to 2016 time period. 

While the models presented here are also used for a TSI reconstruction in \citet{Yeo:etal:2020:Science}, we focus in this publication on analyzing their
radiative properties directly.  Our approach allows to quantify the contribution of small-scale magnetic field, much of which is hidden in current observations 
\citep[e.g.][]{Danilovic:2016:SSD_Power,DelPinoAleman:2018:SSD_Hanle},  to solar irradiance. In particular
 we compute the TSI and SSI sensitivity of the quiet Sun, i.e. by how much TSI and SSI would change if the field strength would vary around the currently known
 nominal quiet Sun level. This approach does not present an ab initio model for a grand minimum, but it does provide physics based constraints on possible 
 scenarios as it relates irradiance changes directly to the required changes in the mean photospheric magnetic field strength.

\begin{figure*}
        \centering
        \resizebox{0.95\hsize}{!}{\includegraphics{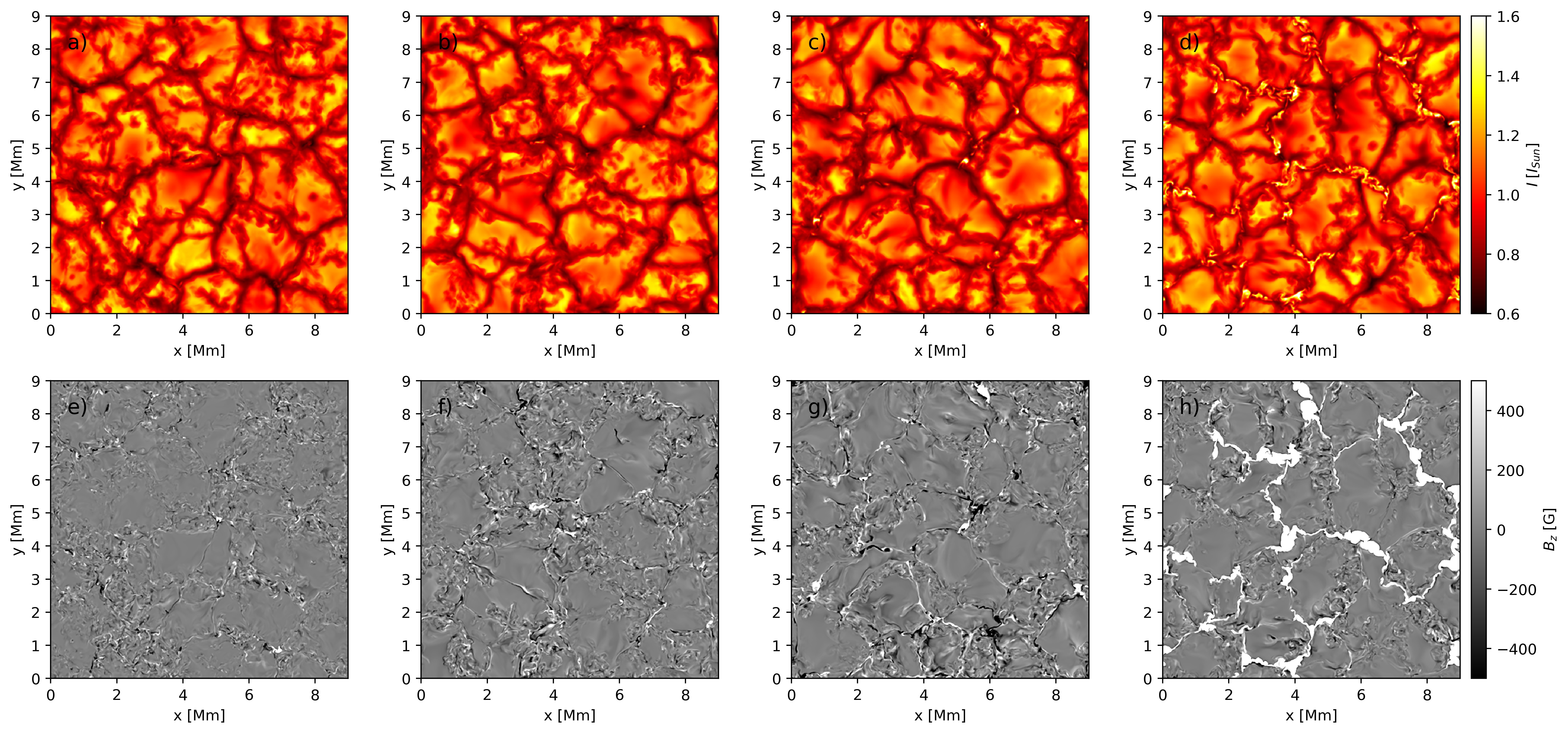}}
        \caption{Top: Bolometric intensity for a vertical ray direction for the setups (left to right) SSD$_{\rm Weak}$, SSD$_{\rm QS}$, SSD$_{\rm Strong}$, and N100. Bottom: Corresponding 
        	vertical magnetic field on $\tau=1$ surface.}
        \label{fig:1}
\end{figure*}

\section{Simulation setup}

\subsection{Small-Scale Dynamo simulations}
In this work we use the MURaM radiative MHD code that was originally developed by \citet{Voegler:etal:2005}. In particular we perform small-scale dynamo simulation based on 
the setup of \citet{Rempel:2014:SSD}. In their work it was found that the saturation level of the dynamo
is dependent on details of the lower domain boundary condition that essentially parameterizes the coupling to the deeper parts of the solar convection zone.
Simulations that minimize this coupling (open for flows, but no transport of magnetic field into to the domain) saturate at a level that is about a factor of 2
lower than current inferences from Zeeman and Hanle measurements \citep{Danilovic:2016:SSD_Power,DelPinoAleman:2018:SSD_Hanle}. The resulting photospheric magnetic field is mostly 
a consequence of shallow recirculation and is organized on small scales \citep{Rempel:2018:expl}. Allowing for deep recirculation through the presence of horizontal field
in upflow regions at the lower boundary condition leads to a saturation field strength more in line with Zeeman and Hanle measurements and adds in addition a more
organized component to the photospheric magnetic field. As analyzed in \citet{Rempel:2018:expl}, horizontal expansion due to stratification significantly smoothes magnetic field
transported into the photosphere from deeper layers in the center of granules. The subsequent amplification through concentration into photospheric downflow lanes
leads to thin sheets reaching kG strength and having a horizontal extent comparable to the size of granules. 

On the one hand, this strong dependence on the lower 
boundary condition makes it impossible to construct fully self-consistent models of quiet Sun magnetism in local domains that focus with high resolution on the photospheric
layers alone. Even the small-scale dynamo is distributed across the convection zone and contributions from the deeper regions of the convection zone matter. 

On the 
other hand, we can easily take advantage of this boundary condition dependence to create quiet Sun models with different levels of magnetization in order to study the effect
of small-scale photospheric field on spectral solar irradiance. To this end we focus in this investigation on 5 setups. They all have in common a simulation domain with 
$9$~Mm horizontal extent (periodic) and a vertical extent of $3.2$~Mm (the photosphere is located about 1 Mm beneath the top boundary). Horizontal/vertical grid spacing
is $17.58$/ $10$ km. The 5 setups differ in terms of their magnetic properties through initial state and bottom boundary conditions:
\begin{enumerate}
\item[(1)] a non-magnetic reference, we refer to this model in the following as HD.
\item[(2)] a model representing a weak quiet Sun. Here we use a vertical field magnetic field lower boundary condition similar to \citet{Voegler:Schuessler:2007}. This model
has no contributions from the deeper convection zone and represents an activity level that can be maintained by a small-scale dynamo operating just in the upper-most 2Mm
of the convection zone through shallow recirculation. We find a mean vertical magnetic field strength of about $\umz=45$~G in the photosphere and refer to this model 
in the following as SSD$_{\rm Weak}$
\item[(3)] a model representing a current day quiet Sun. Following \citet{Rempel:2014:SSD} we impose a symmetric boundary condition on all 3 magnetic field components, i.e. 
we allow for the presence of horizontal field in inflow regions. The horizontal RMS field strength in inflow regions at the lower boundary is limited to $300$~G. This setup
leads on average to a vertical magnetic field strength of $\umz=69$~G  in the photosphere and is referred to as SSD$_{\rm QS}$ in the following. 
\item[(4)] a model representing a more active quiet Sun. The setup is similar to SSD$_{\rm QS}$, but we did not impose the $300$~G limit on the horizontal RMS field strength in 
inflow regions. This setup leads on average to about $\umz=85$~G vertical magnetic field strength  in the photosphere and is referred to as SSD$_{\rm Strong}$ in the following. 
\item[(5)] a model representing a weak 'network'. This model uses the same boundary conditions as setup (3), but starts from an initial state 
with uniform $100$~G vertical magnetic field. The model reaches a vertical magnetic field strength  of about $\umz=156$~G and is referred to as N100 in the following.
\end{enumerate}
The cases (2) - (4) have zero vertical magnetic netflux and were initialized with a $10^{-3}$~G vertical random magnetic field, the case (5) started initially with a uniform 100 G 
vertical magnetic field. Figure \ref{fig:1} shows vertical bolometric intensity as well as $B_z(\tau=1)$ (magnetogram) snapshots for the setups (2) - (5) considered in this work.
We use in our investigation the N100 setup in order to explore the impact of network field on TSI and SSI, however, the N100 setup on its own is not considered a representation of 
quiet Sun.

All simulations use  identical boundary conditions for the hydrodynamic variables (as described in \citet{Rempel:2014:SSD}, boundary "OSb"). At the bottom boundary all three 
components of the massflux are symmetric across the boundary. The mean gas pressure is fixed at the boundary and extrapolated hydrostatically into the boundary layers, while (damped) 
pressure fluctuations are allowed. The entropy is set to a fixed value in inflow regions, whereas outflows carry plasma with their entropy out of the domain (symmetric condition on entropy).
It is important to note that these boundary conditions do not prescribe a convective energy flux through the bottom of the domain, since both, the entropy contrast between up- and downflows
and convective flow velocities can freely adjust. In the saturated state the energy flux is determined by radiative losses in the 
photosphere, which depend on (1) the entropy of fluid that enters in inflow regions at the bottom boundary and is transported into the photosphere, and (2) magnetic field present in the photosphere.
The convective energy flux throughout the domain adjusts (together with the mean stratification) to compensate the radiative losses in the photosphere. 
We have chosen an inflow entropy value that gives an average energy flux of $6.3\times 10^{10}$ erg cm$^{-2}$ s$^{-1}$ for the case SSD$_{\rm QS}$. This entropy value is used in all other setups, which 
essentially assumes that thermodynamic properties of the deeper convection zone do not change in response to the different magnetic conditions we study. Through a fixed inflow entropy across 
the series of setups studied here, we limit the investigation to irradiance changes that result solely from the influence of magnetic field present in the photosphere. This assumption is justified for
times scales that are short compared to the Kelvin Helmholtz time scale of the convection zone, which is about $200,000$ years.

\subsection{Spectral solar irradiance computation}
In this investigation we are primarily interested in spectral solar irradiance (SSI) and to a lesser degree in the detailed radiative properties of individual magnetic features.
Since computing intensity maps for different frequencies and inclination angles as a post-processing step can be very time consuming and data intensive, we expanded the
radiative transfer (RT)  scheme of the MURaM code to allow for additional diagnostic RT calculations in addition to the standard non-grey RT that is performed
every time-step to compute the radiative heating/cooling required for evolving the MHD system. While the non-grey RT is computed with 4 opacity bins every time-step,
we perform the diagnostic RT with about 200 wavelength bins every 50 time-steps, leading overall to about a doubling of the computing time requirements for RT. From the diagnostic
RT we save among other quantities the outgoing spectral energy flux, which we will use in the following analysis. Since our simulations target only quiet Sun conditions, our simulation
domain is representative for the solar photosphere on all latitudes and longitudes, i.e. the outgoing energy flux from the local Cartesian simulation domain can be directly converted
into irradiance assuming spherical symmetry. 

\begin{figure*}
        \centering
        \resizebox{0.95\hsize}{!}{\includegraphics{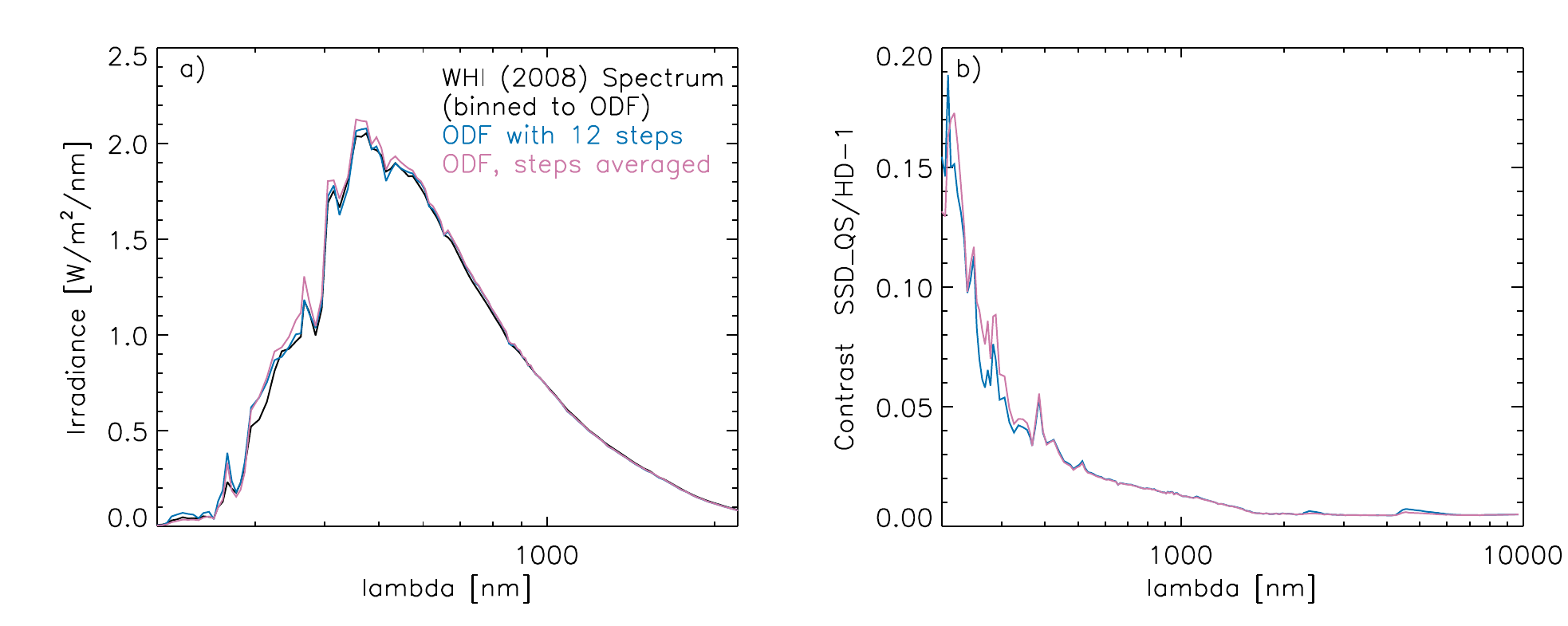}}
        \caption{Panel (a): Comparison of computed spectra (ODF$_{12}$, blue) and (ODF$_{\rm MEAN}$, magenta) to an observed spectrum (black). The latter is binned to the computed wavelength 
        	intervals for better comparability. Panel (b): Relative contrast between a snapshot from setups (1) and (3). While the opacity binning approaches underlying both spectra
	differ in absolute terms in certain spectral ranges by a few $\%$, both approaches capture irradiance contrasts with sufficient accuracy when used in a relative manner.}
        \label{fig:2}
\end{figure*}

\subsection{Opacity source}
We base our computation on opacity distributions functions (ODFs) provided by \citet{Castelli:Kurucz:2004} \footnote{http://wwwuser.oats.inaf.it/castelli/odfnew.html}. 
In particular we use the low resolution ODF with 328 wavelength intervals
computed for solar metallicity and 2~km~s$^{-1}$ micro-turbulence. For consistency we computed both
the opacity bins used by MURaM and the bins we use for the diagnostic RT using the same ODF. While the opacity bins are based on the full spectral range, we limit the diagnostic
RT to wavelength larger than 200 nm, which results in about 200 wavelength intervals covering the range from 200 to 10,000 nm with spacings from about 5 nm at the shortest and 
200 nm at the longest wavelength. Each wavelength interval of the ODF is subdivided into 12 opacity steps. In order to limit the computational effort for the diagnostic RT we represent each
wavelength interval by a single opacity that is computed through a harmonic mean (since all opacity steps within a wavelength interval have no wavelength dependence, the Rosseland mean
reduces to a harmonic mean). Since the opacity steps do have individual weights $w_i$ ($\sum_{i=1}^{12} w_i =1$), the resulting mean opacity is given by
\begin{equation}
\frac{1}{\bar{\kappa}}=\sum_{i=1}^{12} \frac{w_i}{\kappa_i}\label{eq:1}
\end{equation}
In the following we refer with ODF$_{12}$ to the approach that explicitly computes all 12 opacity steps and with ODF$_{\rm MEAN}$ to the approach that uses the harmonic opacity mean.
Figure \ref{fig:2} compares spectra with ODF$_{12}$ and ODF$_{\rm MEAN}$ to an observed reference spectrum. In panel (a) the black line shows the observed WHI spectrum from 2008 
\citet{Woods:etal:2009:WHI} that has been binned
to the wavelength intervals of the computed ODF based spectrum for better comparability. The blue line shows the ODF$_{12}$ spectrum, whereas the magenta line uses ODF$_{\rm MEAN}$.
The two approaches lead to differences in TSI of about $1\%$, but differences can be substantially larger for certain wavelengths shorter than $400$~nm. The harmonic mean of the ODF opacity steps 
in ODF$_{\rm MEAN}$ leads to a systematic overestimation of the outgoing energy flux. For the analysis in this paper we are however interested in relative changes between the setups (1) - (5) and to this 
end TSI and SSI will be computed in relative terms. This is illustrated in panel (b). Here we take a snapshot from setups (1) and (3) and compute the SSI contrast by using either ODF$_{12}$ or
ODF$_{\rm MEAN}$ for both setups.  Although both approaches differ in absolute terms by more than $1\%$, contrasts are still well captured at levels as low as $0.01\%$. Since using ODF$_{\rm MEAN}$ 
is 12 times less expensive as ODF$_{12}$, all computations in the following will be based on that opacity source. We refer to \citet{Criscuoli:etal:2020:irrad_codes} for a more detailed discussion on how
spectral solar irradiance computations are dependent on details of radiative transfer codes, opacity sources and LTE vs. NLTE effects.

\section{Results}

\begin{figure*}
        \centering
        \resizebox{0.95\hsize}{!}{\includegraphics{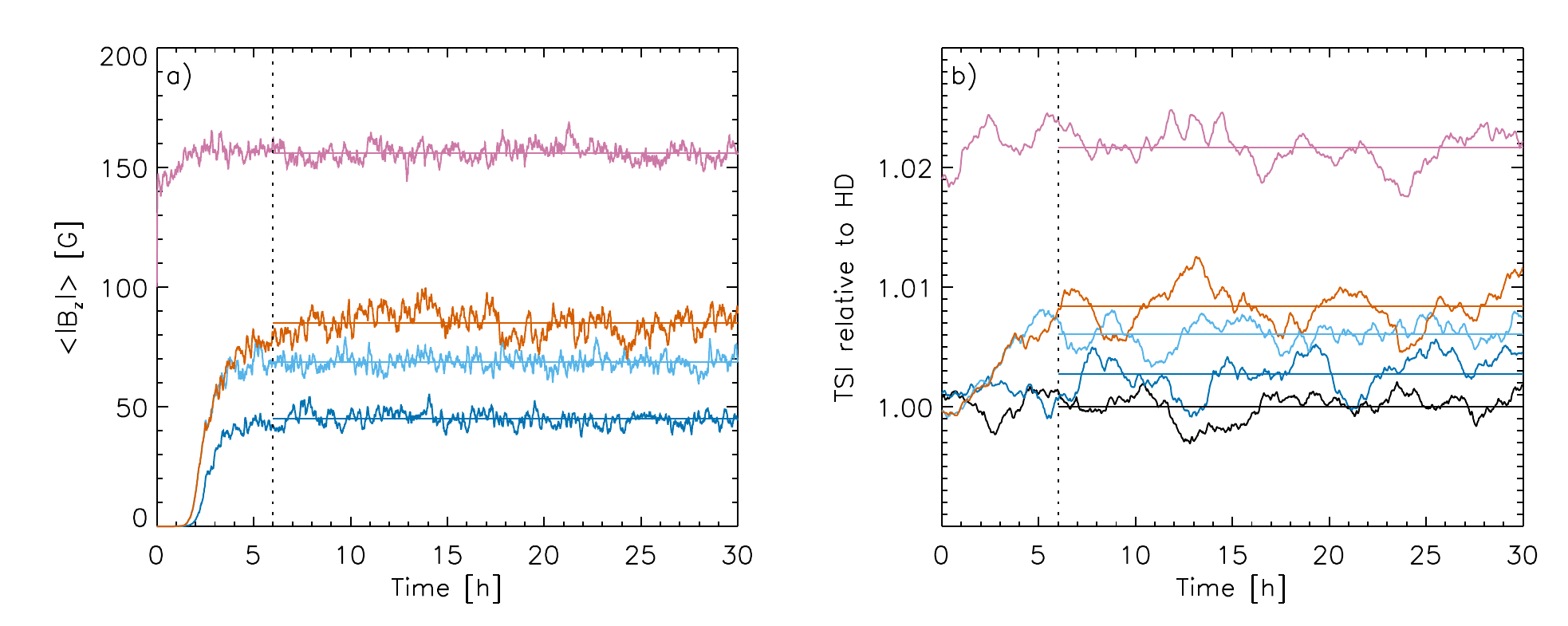}}
        \caption{Panel (a): Time evolution of the mean vertical magnetic field strength $\umz$ for the setups SSD$_{\rm WEAK}$ (dark blue), SSD$_{\rm QS}$ (light blue), SSD$_{\rm Strong}$ (orange), and N100 (magenta).
       	All cases were started at $t=0$~hours from the same HD snapshot (with different magnetic initial states). Panel (b): Time evolution of the outgoing radiative energy flux for the setups (1) - (5). The HD case is shown in black. For
	the following analysis we consider $24$ hour averages as indicated in the figure. In the right panel we apply a 2 hour running boxcar average for better readability, since short-term TSI fluctuations have amplitudes of $1-2\%$.}
        \label{fig:3}
\end{figure*}

\begin{figure*}
        \centering
        \resizebox{0.95\hsize}{!}{\includegraphics{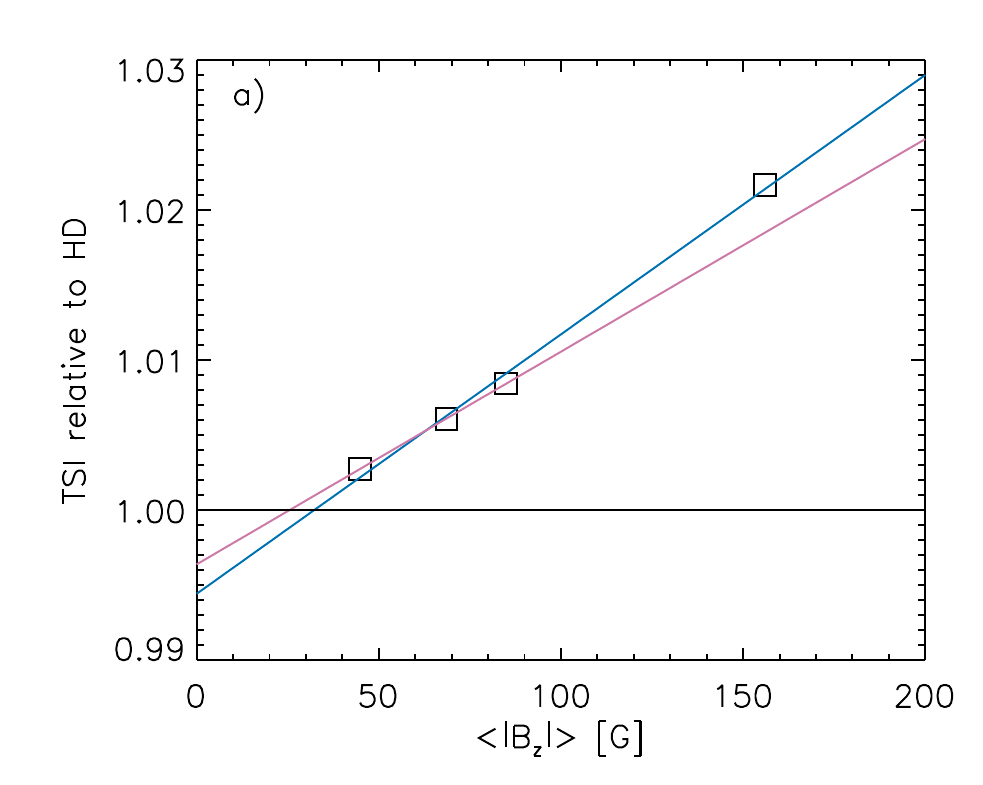}\includegraphics{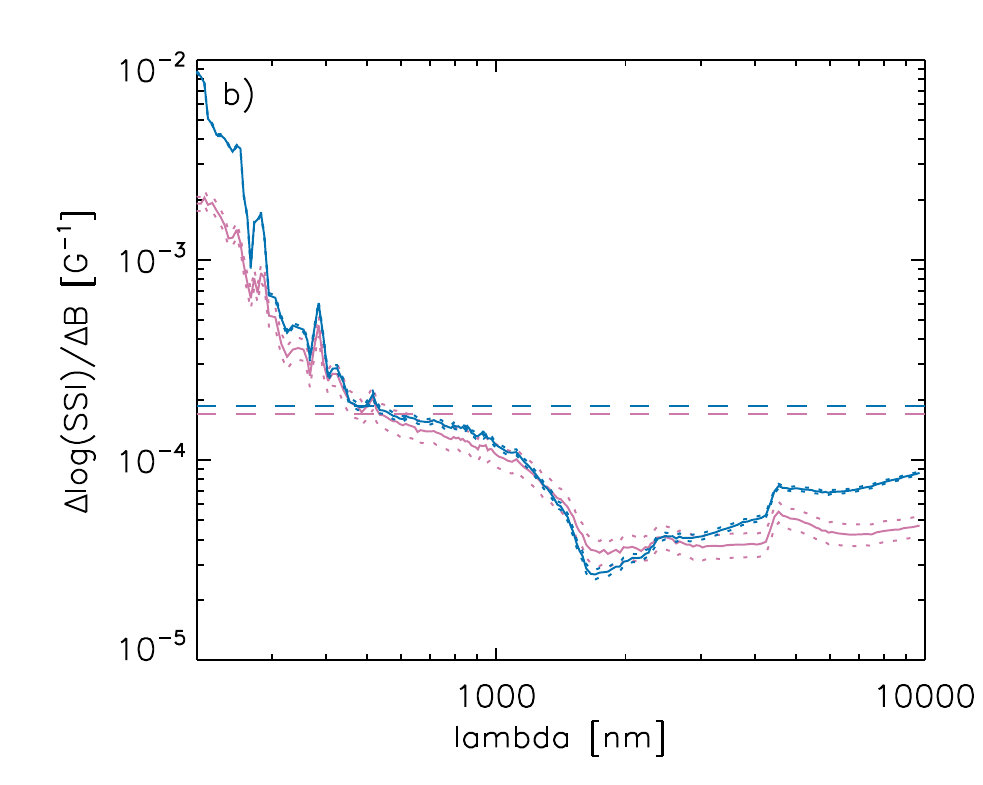}}
        \caption{Panel (a): Variation of TSI with the mean vertical field strength $\umz$ in the photosphere. The squares show the data points from Table \ref{tab:1}. The magenta line is a linear regression using only the SSD data points,
        the blue line also considers the N100 case. The regression slopes and their error are displayed in Table \ref{tab:2}. Panel (b): Sensitivity of SSI in the range from 200 to 10,000 nm. The solid magenta line shows the 
        	wavelength dependent regression slope based on the SSD setups, the solid blue line based on the SSD and N100 setups. Dotted lines indicate the formal error bars, dashed lines the TSI based regression 
	slopes from Table \ref{tab:2}. }
        \label{fig:4}
\end{figure*}

\begin{figure*}
        \centering
        \resizebox{0.95\hsize}{!}{\includegraphics{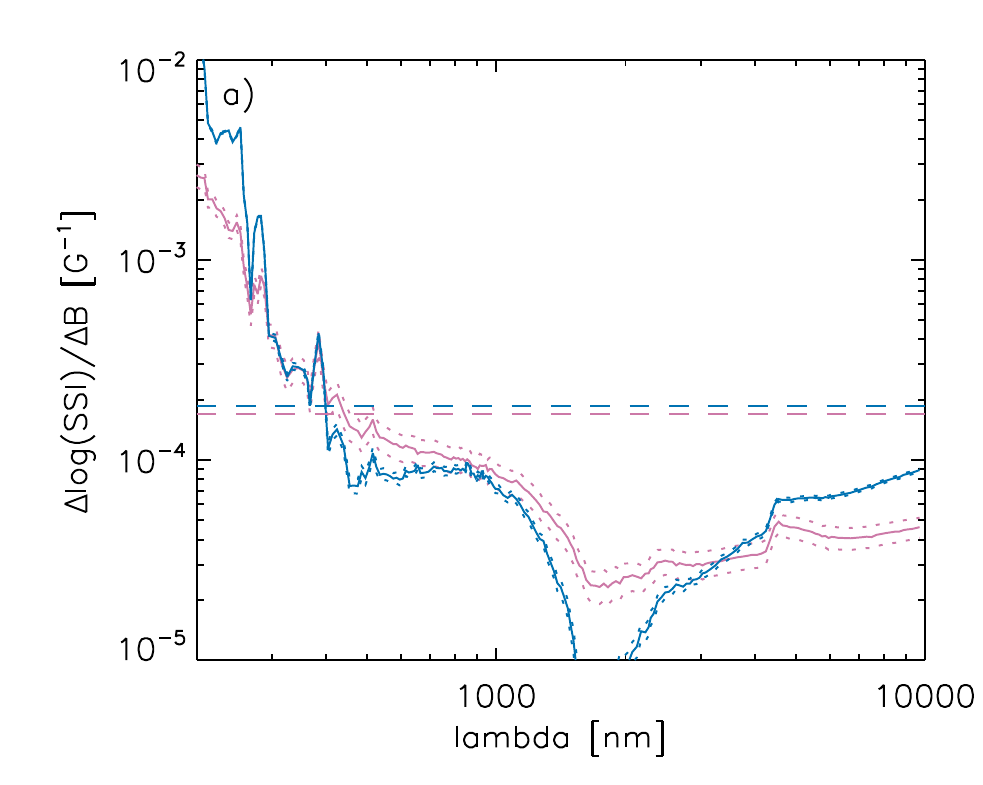}\includegraphics{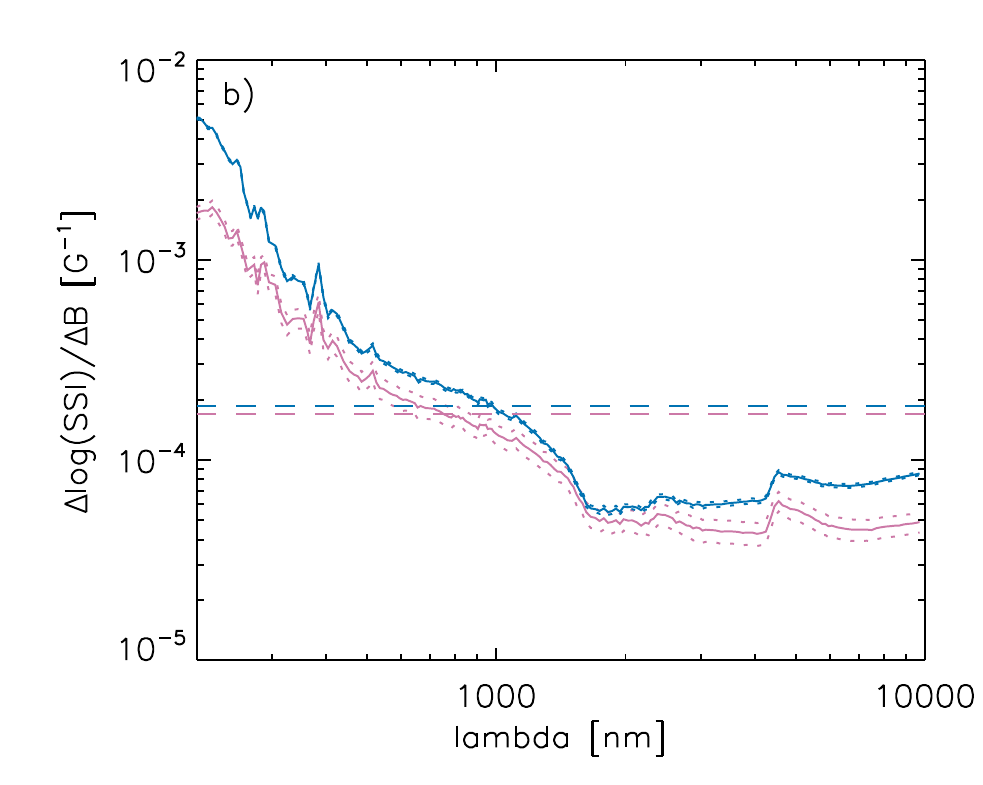}}
        \caption{Panel (a): Sensitivity of SSI in the range from 200 to 10,000 nm for a vertical ray; (b) for an inclined ray with $\mu=\cos(\theta)=0.33$. Dotted lines indicate the formal error bars, dashed lines the TSI based regression 
	slopes from Table \ref{tab:2}. The red line is a linear regression using only the SSD data points, the blue line also considers the N100 case.}
        \label{fig:5}
\end{figure*}

\subsection{Temporal evolution of magnetic field and TSI}

\begin{table}[]
\centering
\caption{Mean vertical field strength and TSI relative to HD}
\label{tab:1}
\begin{tabular}{l|c|c}
Setup  & $\umz$ [G] & $\Delta$ TSI [$\%$]  \\
\hline 
SSD$_{\rm WEAK}$  & 44.87 $\pm$ 0.39 & 0.275  $\pm$ 0.047\\
SSD$_{\rm QS}$  & 68.61 $\pm$ 0.32&  0.608  $\pm$ 0.036\\
SSD$_{\rm Strong}$  & 84.95  $\pm$ 1.56& 0.842 $\pm$ 0.052\\
N100 & 156.05 $\pm$ 0.48& 2.17 $\pm$ 0.043
\end{tabular}
\end{table}

\begin{table}[]
\centering
\caption{Regression slopes}
\label{tab:2}
\begin{tabular}{l|c|c}
Setup  &  $\Delta\log({\rm TSI})/\Delta B$  [G$^{-1}$] & slope error [G$^{-1}$] \\
\hline 
SSD & $1.42\times 10^{-4}$ &  $0.18\times 10^{-4} $\\
SSD + N100& $1.73 \times 10^{-4}$ &  $ 0.053\times 10^{-4}$

\end{tabular}
\end{table}

Figure \ref{fig:3} shows the time evolution of the setups (1) - (5) in terms of their photospheric field strength (panel a) and outgoing energy flux (equivalent to TSI) normalized to the HD case (panel b).
All setups were started at $t=0$ hours from a thermally relaxed HD simulation. The 3 setups without a net flux imbalance where initialized with a $10^{-3}$~G random vertical magnetic field and evolved for a total
of $30$ hours. Initially the weak seed field is amplified by a small-scale dynamo, until a saturated regime is reached after about $6$ hours. The network setup was initialized at  $t=0$ hours with a uniform vertical
magnetic field of $100$~G and evolved for another $30$ hours. In this case a saturated state is reached after about $2$ hours. In all cases we use the last $24$ hours to compute averages of mean vertical 
magnetic field and TSI, which are provided in Table \ref{tab:1}. The TSI shows a substantial amount of fluctuation, mostly due to granulation noise. In panel (b) we already applied a 2 hour running boxcar 
average to suppress the short-term variation. The minimum to maximum variation is on the order of a few  $\%$ of the mean energy flux, the $1\sigma$ fluctuation is about $0.55\%$, which is still comparable 
to the difference between the HD and SSD$_{\rm QS}$ setups. Because of that it is crucial to average over a large enough number of independent realizations of granulation in order to obtain sufficiently
well defined average quantities. This can be either achieved by performing simulations with a large horizontal domain extent, or by evolving a small domain for a sufficiently long time interval, in this
study we do the latter. We base the following analysis on quantities that have been averaged over 24 hours of simulated time. The $9\times 9$~Mm$^2$ horizontal domain extent provides about $30$
granules at any given time. Assuming a typical granular life-time of $10$ minutes, $24$ hours provide an average over about $4000$ independent granules, which is comparable to what is found in 
a $100\times 100$~Mm$^2$ patch of granulation at any given time. The corresponding error for the $24$ hour TSI average is about $0.04\%$ (with a granulation life-time of about 10 minutes, there are about 
144 independent realizations of granulation in a $24$ hour time series). This crude error estimate is consistent with a more formal approach outlined in Appendix \ref{App:A} that considers the covariance of 
the time series. The resulting error estimates are in the $0.036 \%- 0.053\%$ range as shown in Table \ref{tab:1} and are further used in the following analysis. 

\subsection{TSI and SSI sensitivity to photospheric field strength}
In Figure \ref{fig:4}(a) we plot the $24$ hour averaged TSI as function of the $24$ hour averaged mean vertical magnetic field strength (squares). The magenta and blue lines  show a linear regression based on only the dynamo cases
without netflux (magenta) and the dynamo cases including the $100$~G network case (blue). The corresponding regression slopes and their formal errors due to granulation noise are shown in Table \ref{tab:2}. 
We treat the errors in magnetic field strength and TSI as uncorrelated. This provides a conservative estimate, since a weak positive correlation is present. 
The error of the slope based on the SSD cases
is larger due to the small separation in terms of photospheric field strength, compared to the regression also including the N100 setup. We find that the regression slope including the network case is about $20\%$ steeper, however, that difference
lies just barely outside the error intervals.  The linear regression does not reach 
the TSI value of the HD setup at $0$~G field strength, but rather for a value of about $25-30$~G. This is consistent with the finding of \citet{Rempel:2014:SSD} that kG flux concentrations only form in the photosphere for a
mean vertical field strength exceeding a value of  $\umz\approx30$~G. For setups  in excess of $30$~G the fill factor of kG field was found to increase more or less linearly with field strength, which is
reflected in the linear increase of TSI.  

Figure \ref{fig:4}(b) shows the corresponding regression slopes for the spectral solar irradiance (SSI). Similar to Figure \ref{fig:4} we show here 2 slopes, one based only on the SSD setups (magenta) and one based on the SSD + N100
setups (blue). Both slopes are similar within error bars in the  $300$-$1200$ nm range, while clear differences exist for shorter and longer wavelength. The slope based on the SSD+N100 setups shows more sensitivity
for wavelength of less than $350$~nm and more than $3000$~nm and reduced sensitivity in the $1200$ to $3000$~nm range. The rather similar slopes based on the TSI are a result of the similar SSI slopes in the  $300$-$1200$
interval that contributes about $77\%$ of the solar energy flux and a partial compensation of differences in the other wavelength ranges. The differences at short and long wavelength arise from higher layers in the photosphere, 
where the N100 case has a substantially different field strength and structure. Furthermore, even the similarity of the SSI sensitivity in the $300$-$1200$ nm range is mostly coincidental as shown in Figure \ref{fig:5}. For a vertical ray the sensitivity
derived from the SSD+N100 cases is systematically lower in the 400-3000 nm range compared to the SSD cases alone, Figure \ref{fig:5}(a), whereas the opposite is true for an inclined ray with $\mu=\cos(\theta)=0.33$, Figure \ref{fig:5}(b). The N100 case does have magnetic
flux concentrations that are systematically larger than those found in the SSD cases \citep[see][for a more detailed analysis]{Peck:etal:2019:structure_size}. While they appear darker for a vertical ray, the intensity of inclined rays is 
enhanced due to a more pronounced 'hot wall' effect. Interestingly many of these differences in details  compensate each other and lead to a similar TSI sensitivity as shown in Figure \ref{fig:4}(a).  

\begin{figure*}
        \centering
        \resizebox{0.95\hsize}{!}{\includegraphics{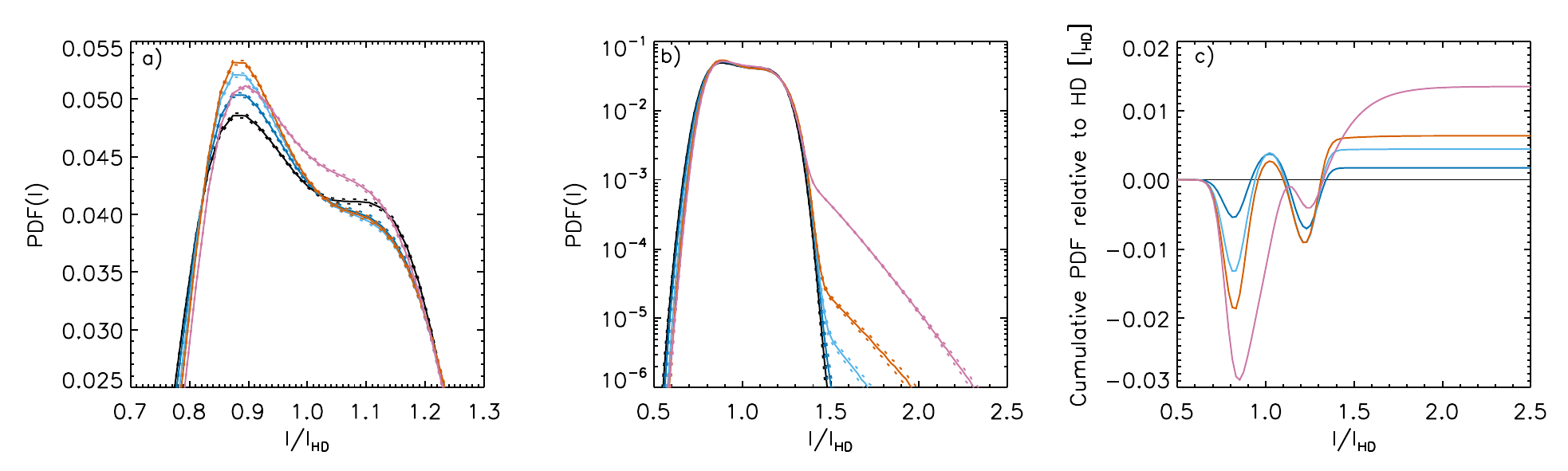}}
   	 \caption{Probability density function (PDF) of the bolometric intensity (vertical ray direction) for the setups HD (black), SSD$_{\rm WEAK}$ (dark blue), SSD$_{\rm QS}$ (light blue), SSD$_{\rm Strong}$ (orange), and N100 (magenta).
       	Panel a) presents the core of the PDF on a linear scale, panel b) presents the wings on a logarithmic scale. Dotted lines indicate the formal error intervals from granulation noise. Panel c) shows the cumulative
	intensity PDFs relative to the HD case. }
        \label{fig:6}
\end{figure*}

\begin{figure*}
        \centering
        \resizebox{0.95\hsize}{!}{\includegraphics{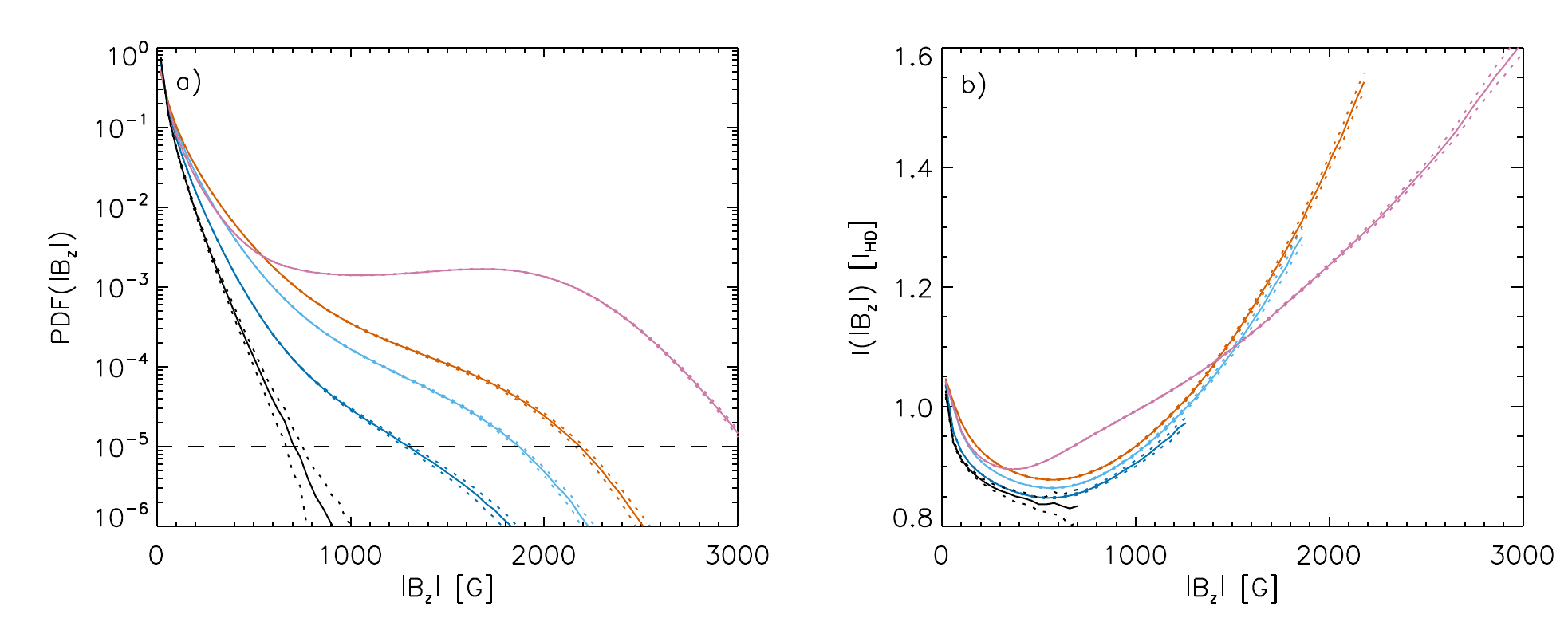}}
        \caption{a) Probability density function (PDF) of the vertical magnetic field strength for the setups SSD$_{\rm WEAK}$ (dark blue), SSD$_{\rm QS}$ (light blue), SSD$_{\rm Strong}$ (orange), and N100 (magenta).
        In addition the black line presents the PDF for an interval during the kinematic growth phase of the dynamo where 25~G $< \langle\vert B_z\vert\rangle <$ 35~G. b) Relation between 
        vertical field strength and bolometric intensity (vertical ray). Dotted lines indicate the formal error intervals from granulation noise.
        }
        \label{fig:7}
\end{figure*}

\subsection{Intensity and vertical magnetic field strength histograms}
\label{I_B_histo}
In this section we analyze how the presence of magnetic field changes the radiative properties of the photosphere by looking at distribution function of  intensity, vertical field strength as well as their relation. We restrict the analysis to the intensity
of the vertical ray direction in order to allow for a pixel-by-pixel comparison with the vertical magnetic field strength at $\tau=1$, but also note that there are distinct differences for inclined rays as shown in Figure \ref{fig:5} that are not captured here.
Furthermore we want to stress that many of the changes discussed here are subtle and require a large number of snapshots to be statistically significant. We use for each setup about 5000 snapshots spanning $24$ hours of time evolution, 
the resulting formal errors from granulation noise are indicated in all plots. The indicated  error level is $\sigma/\sqrt{144}$ (assuming a $10$ minute lifetime for the granulation pattern).

Figure \ref{fig:6} shows the probability density function (PDF) of the bolometric intensity for a vertical ray direction. The line color corresponds to the HD (black), SSD$_{\rm WEAK}$ (dark blue), SSD$_{\rm QS}$ (light blue), SSD$_{\rm Strong}$ (orange), and N100 (magenta)
setups.  Panel a) highlights changes in the core of the distribution, while panel b) highlights the wings of the distribution.  It is evident that even the quiet Sun magnetic field leads to significant changes of the intensity PDF for all intensity values. In the
core we find an increase in the asymmetry compared to the HD case with increasing SSD field strength. Compared to the HD case, the peak at $0.9\,I_{\rm Sun}$ increases, whereas the plateau at $1.1\,I_{\rm Sun}$ is reduced. 
In the wings we find a reduction of pixels with low intensity and the development of a tail with high intensity values. While the latter is a reflection of the fact that the dark downflow lanes are filled with bright magnetic flux concentrations, changes to 
the core of the PDF are more subtle and are in part related to the effect that the SSD magnetic field suppresses small-scale turbulence at the edge of granules. Here we also find distinct difference in the shape of the PDF for the N100 setups compared to the
SSD setups. 
Changes to the PDF at all intensity values do make significant contributions to the overall intensity change as highlighted in panel c). Here we present the cumulative PDFs of all the magnetic cases relative to the HD case. Interestingly, most contributions from
pixels with $I< 1.3 I_{\rm Sun}$ do balance out, such that the effective brightness change is similar to the contributions from bright pixels with  $I> 1.3 I_{\rm Sun}$ alone.

Figure \ref{fig:7}a) shows the probability density function of the vertical magnetic field strength at $\tau=1$ for the setups SSD$_{\rm WEAK}$ (dark blue), SSD$_{\rm QS}$ (light blue), SSD$_{\rm Strong}$ (orange), and N100 (magenta). In addition the black
line refers to the PDF found in the photosphere for a field strength of $\langle\vert B_z\vert\rangle_{\tau=1} \approx 30$~G, which corresponds to the field strength at which we found in Figure \ref{fig:4}a) the intersection of the regression curve with the
TSI value corresponding to the HD case. We computed the PDF for this case by averaging snapshots with $\langle\vert B_z\vert\rangle_{\tau=1} $ from 25 to 35~G during the kinematic growth phase of  SSD$_{\rm WEAK}$. Since this time period lasts 
only about $20$ minutes, we use $\sigma/\sqrt{2}$ to indicate the error from granulation noise. For these snapshots
we do not find values of $\vert B_z\vert_{\tau=1}$ exceeding 1~kG, unlike the other setups that have extended tails of the distribution in the 1-3 kG range.

Figure \ref{fig:7}b) shows the pixel-by-pixel vertical bolometric intensity as function of the vertical magnetic field strength. We restrict here that analysis to a field strength range for which the PDF has values above $10^{-5}$ as indicated in panel a).
 Regions with weak field correspond mostly to granular upflows and have intensities about 5\% larger than the mean.
Stronger magnetic field is found in downflow lanes with lower intensity. The relationship changes for field strength exceeding values of about $600$~G for the SSD setups (about $400$~G for the N100 setups), at which we see a monotonic increase 
of intensity with field strength.
Comparing the SSD cases, we find that they all follow a curve of similar shape, for stronger field the curve is mostly extended towards higher field strength values. There is however also a secondary effect that leads to an upward shift of the
$I(\vert B_z\vert)$ relation that affects even the pixels with weaker magnetic field strength. This can be explained primarily as a "displacement" effect. In stronger SSD setups the cores of the downflow lanes are filled with the strongest
fields, which displaces the weaker fields more towards the edge of downflows that do have higher intensity values. This does not mean that weaker magnetic fields have a stronger back-reaction that leads to intensity changes in the stronger
SSD cases. We also show the $I(\vert B_z\vert)$ relation for the photosphere with $\langle\vert B_z\vert\rangle_{\tau=1} \approx 30$~G. In this case we do not see any enhancement of brightness with increasing field strength,
which suggests that the magnetic field is too weak to have any feedback that could change the radiative properties of the photosphere, i.e. the magnetic field is mostly a passive tracer that populates the downflow lanes, with the strongest field
typically found in the strongest and darkest downflow regions. This explains why the TSI regression in Figure \ref{fig:4}a) reaches the HD value for a strength of $30$~G.

The $I(\vert B_z\vert)$ for the N100 setup has a qualitatively similar shape to the SSD setups, however, pixels with $\vert B_z\vert< 1.5$~kG are generally brighter and pixels with $\vert B_z\vert> 1.5$~kG are darker compared to the SSD setups. The brightening
of the weaker field pixels can be explained again by the displacement effect described above, the darkening of the strong field pixels is caused by the organization into larger flux concentrations in the case of the network setup as it has been 
demonstrated by  \citet{Peck:etal:2019:structure_size}. By looking at identifiable flux concentrations (which is different from a pixel-by-pixel comparison) they found that the brightness of magnetic elements depends on their size and that that
relation is (within error bars) universal in SSD and network cases. What changes with the magnetic environment is the size distribution of structures. They found in network cases more larger structures with a lower intensity contrast.

We refer to Appendix \ref{App:B} for further discussion on how the properties presented here for the bolometric intensity depend on wavelength.

\begin{figure*}
        \centering
        \resizebox{0.95\hsize}{!}{\includegraphics{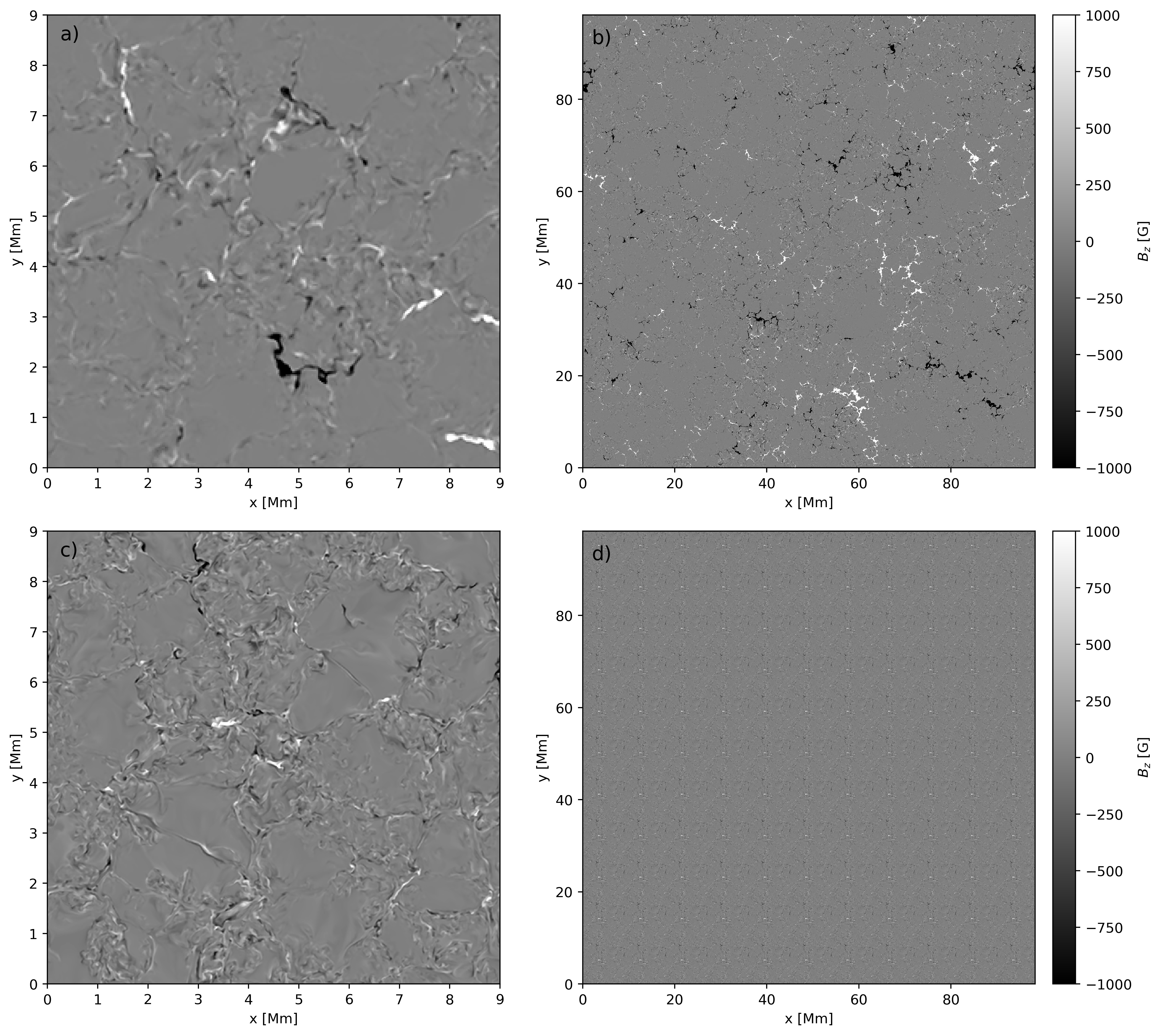}}
        \caption{Comparison of $\umz$ for small-scale dynamo simulations in small and large domains. Panels (a,b) show a dynamo simulation in a 98.304 Mm wide and 18.432 Mm deep domain from \citet{Rempel:2014:SSD}. 
       	 	Panel (a) shows a 9 Mm wide subsection, panel (b) the full horizontal domain extent. Panels (c,d) show for comparison a periodic replication of the 9x9 Mm$^2$ wide SSD$_{\rm QS}$ in a 9 and 98.304 Mm wide domain.
		While both presented simulations have zero netflux ($\smz=0$) and an average vertical field strength of about $\umz\approx 67$ G, the organization of magnetic field on larger scales is substantially different.}
        \label{fig:8}
\end{figure*}

\subsection{Quiet Sun Network}
The SSD setups in only $9$~Mm wide domains do underestimate flux imbalance on meso- to super-granular scales. By construction the netflux imbalance on the scale of the simulation domain is zero, which is not necessarily the case on a comparable scale in the
quiet Sun. In order to illustrate this shortcoming we present in Figure \ref{fig:8} the simulation run O32bSG from \citet{Rempel:2014:SSD} and compare it to SSD$_{\rm QS}$ from this investigation.  O32bSG was computed in a $98.304$ Mm 
wide and $18.432$ Mm deep 
domain with an isotropic grid spacing of $32$~km. The numerical setup is otherwise similar to SSD$_{\rm QS}$ (apart from using grey RT and not imposing a limit to the horizontal field strength in 18 Mm depth). In panel (a) we present a 
$9\times 9$~Mm$^2$ cutout of O32bSG, in panel (b) the full horizontal extent. Panel (c) shows the $9\times 9$~Mm$^2$ domain of SSD$_{\rm QS}$, for comparison to O32bSG we replicate in panel (d) SSD$_{\rm QS}$ to fill the 
full domain extent of O32bSG. We note that both simulations have zero netflux (on the respective extent of the simulation domain) and a comparable mean vertical magnetic field strength ($\umz\approx 67$~G) for the snapshots shown in Figure \ref{fig:8}. On small scales differences are moderate, mostly caused by the different
grid spacing in both simulations. O32bSG shows indications of a few stronger flux concentrations than SSD$_{\rm QS}$. The differences on large scales are striking. O32bSG shows a mixed polarity quiet Sun network on meso- to super-granular scales,
which is not captured in SSD$_{\rm QS}$ by construction. In O32bSG we find that the typical flux imbalance on the scale of $9$~Mm is about $15$~G, on the scale of 27 Mm the imbalance is about half of that. These values agree at least on a qualitative level
with observations. Using the quiet Sun Hinode map studied in detail by \citet{Lites:etal:2008} we find about 7 G flux imbalance on a scale of 9 Mm, 5 G on the scale 27 Mm, 2.5 G on the scale of the entire Hinode map. 

This comparison highlights that perhaps SSD$_{\rm QS}$ with an added flux imbalance on the order of 5-10 G would be a better reference for the magnetic field distribution in the quiet Sun. Based on the findings in this study, this detail has
however only minor impact on the TSI and SSI sensitivity. We found that the TSI sensitivity of the SSD cases and SSD + N100 cases are comparable within error margins and that the relevant quantity that influences irradiance is the mean vertical magnetic field 
strength ($\umz$) regardless of the flux imbalance ($\smz$), at least for the weak flux imbalances considered here. Moderate differences would be expected for the SSI sensitivity, which differs for certain wavelength ranges. We expect that the SSI sensitivity of the quiet Sun falls in-between that derived from the SSD and SSD+N100 setups, although more closer to  SSD than SSD+N100.

\section{Conclusions}
We presented a series of small-scale dynamo simulations with no vertical netflux as well as a setup with an additional $100$~G mean flux density. Computing the TSI from these setups we find a general trend of about
$0.142\%$ (SSD only) to $0.173\%$ (SSD + Network) increase for every $10$~G of mean vertical magnetic field strength on the $\tau=1$ level. A variation of the quiet Sun field strength by about $10\%$ (about 7~G) would
lead to a TSI change comparable to the observed solar cycle minimum to maximum variation. 

In addition we found that the TSI sensitivities of pure small-scale dynamo simulations and weak network simulations having an additional flux imbalance on the scale of the simulation domain are comparable. This suggests that the quiet Sun 
TSI is primarily determined by the amount of
unsigned flux in the photosphere ($\umz$), to which degree that flux is present in form of a small-scale turbulent field or network appears to be of secondary importance (at least for the weak 100G network case considered here).  
This is a non-trivial result, our more detailed analysis looking at the SSI sensitivity for rays of different inclinations as well as intensity distribution functions  and pixel-by-pixels brightness-field strength relations does reveal distinct
differences between cases with and without flux imbalance. However, these differences average out when integrating contributions over all ray directions and wavelength. 

On the one hand, this quite significant TSI sensitivity puts strong indirect constraints on how much the quiet Sun can vary with the solar cycle. In current models most of the observed
TSI variation is explained through a combination spot darkening and facular brightening, while the quiet Sun is assumed to be a steady baseline. If there is any additional quiet Sun magnetic field variability, 
it must be well below  $10\%$ in order to be consistent with this picture. This is in line with direct measurements of quiet Sun field \citep[e.g.,][]{Buehler:2013:cyc_dep,Lites:etal:2014:qs_cyc} that did not find a convincing evidence for a cycle variation.
\citet{Lites:2011:hinode_ssd} also showed that the strength of the inter-network field does not depend on the strength of the surrounding network, which strongly supports an independent origin of the quiet Sun magnetic field through a small-scale
dynamo.  However, \citet{Meunier:2018:QS_variability} did find a significant variability (about 25\%) of the quiet Sun with the solar cycle based on MDI measurements. Using the quiet Sun TSI sensitivity we found ($1.42\cdot 10^{-4}$G$^{-1}$), this would suggest  
a $0.18\%$ TSI variability from  the quiet Sun alone, which is almost twice as strong as the observed variation. Their finding of a rather strong variation is not compatible with the TSI sensitivity determined in this investigation.  

On the other hand, we cannot rule out
from available solar observations that there may be long-term trends  in the quiet Sun reference. The TSI sensitivity quantified in this paper indicates that such trends, even if moderate would make a significant contribution
to the solar irradiance variability. While a $10\%$ drop in magnetic activity would account for a TSI change comparable to the current cycle variability, we can use the HD and SSD$_{\rm WEAK}$ setups to put bounds on 
possible changes. The SSD$_{\rm WEAK}$ scenario describes the level of magnetism that could be maintained by near surface dynamo processes without contributions from the deep convection zone 
\citep{Rempel:2014:SSD,Rempel:2018:expl} with a TSI $0.33\%$ (4.5 W$\,$m$^{-2}$) lower than the SSD$_{\rm QS}$. Removing all magnetic field, although very unrealistic, would imply a drop by another $0.28\%$, i.e. 
$0.61\%$ (8 W$\,$m$^{-2}$) total down from the SSD$_{\rm QS}$ case that matches current constraints on quiet Sun magnetism. While we use here the HD case as the extreme scenario, Figure \ref{fig:4}a) suggests that 
a reduction of the photospheric field strength to 30~G leads to a TSI comparable to the HD case, since the remaining field would be too weak to have an effect on irradiance.

Using a reconstruction model that is based on the MHD simulations presented here, \citet{Yeo:etal:2020:Science} estimated that the TSI cannot drop more than $2.1$ Wm$^{-2}$ compared to a regular cycle minimum. 
This estimate is based on the assumption that the small-scale dynamo generated internetwork field provides a lower bound for the solar magnetic activity. This is a conservative estimate, since also the quiet Sun network 
may be maintained by a small-scale dynamo as we showed in this paper.

While TSI changes in the range suggested by \citet{Shapiro:etal:2011} and \citet{Egorova:etal:2018} cannot be ruled out from first principles based on our investigation, they certainly require substantial changes
in the quiet Sun field strength (about a $50\%$ reduction). Such changes are unrealistic if the quiet Sun magnetic field  originates from a small-scale dynamo that operates independently from the  large-scale dynamo responsible for the solar cycle. We consider 
even the SSD$_{\rm WEAK}$ case unrealistic, since it would imply that the deep convection zone becomes either field free or completely decoupled from the photosphere. Independent from that it is far from clear if changes of the global
heliospheric magnetic field (that would be reflected in the modulation potential derived from cosmogenic isotopes) have a direct connection to the small-scale field present in the quiet Sun. Most of the flux closes already low in the solar atmosphere, only larger scale
flux imbalances lead to magnetic field connecting to the corona or even heliosphere. Those larger scale flux imbalances in the current day quiet Sun are only on the order of a few G, i.e. we found 2.5~G on the scale of a Hinode quiet Sun map. 
This limits their potential contribution to TSI significantly. Using the TSI sensitivity we found in this work, 2.5 G flux imbalance would translate to about a $0.035\%$  in TSI. Furthermore we showed that a quiet Sun network naturally emerges from a 
small-scale dynamo operating in wider and deeper domain, i.e. also this field component is likely maintained by processes that are independent from the solar cycle.  Such a view is also supported by \citet{Judge:Saar:2007}, who studied flat activity 
solar-like stars.

\acknowledgements
This material is based upon work supported by the National Center for Atmospheric Research, which is a major facility sponsored by the National Science Foundation under Cooperative Agreement No. 1852977.
We would like to acknowledge high-performance computing support from Cheyenne (doi:10.5065/D6RX99HX) provided by NCAR's Computational and Information Systems Laboratory, sponsored by the National Science Foundation. This work was partially funded by the NASA LWS grant NNX16AB82G. The author acknowledges the fruitful discussions with the members of the \#335 science team
supported by the International Science Institute (ISSI), Bern. The author thanks Giuliana DeToma, Serena Criscuoli, Kok Leng Yeo, and the anonymous referee for comments on the manuscript. Some of the figures within this paper were produced using IDL colour-blind-friendly colour tables \citep[see][]{pjwright2017}.

\appendix
\section{Error estimate for average of finite length correlated time series}
\label{App:A}
We have a finite length time series $\{x_i\}_{i=1}^n$, for which we estimate the average as $\bar{x}={1\over n}\sum_{i=1}^n x_i$. Since the finite length time series is only a single representation of an infinite ensemble,
$\bar{x}$ differs from the true ensemble average $X=\avr{x_i}=\avr{\bar{x}}$. Consequently $\bar{x}$ has a variance that is given by \citep[see,][]{Flyvbjerg:Petersen:1989}:
\begin{equation}
	\Delta^2(\bar{x})=\avr{\bar{x}^2}-\avr{\bar{x}}^2=\frac{1}{n}\left[\gamma_0+2\sum_{k=1}^{n-1}\left(1-\frac{k}{n}\right)\gamma_k\right]\,,
\end{equation}
where $\gamma_k$ denotes the covariance
\begin{equation}
	\gamma_k=\avr{(x_i-X)(x_{i+k}-X)}\,.
\end{equation}
Following \citet{Flyvbjerg:Petersen:1989}, $\gamma_k$ can be estimated from the finite length time series using
\begin{equation}
	c_k=\frac{1}{n-k}\sum_{k=1}^{n-k}(x_i-\bar{x}) (x_{i+k}-\bar{x})\,,
\end{equation}
which provides a biased estimate for $\gamma_k$. Assuming a finite correlation length $\tau \ll n$ and finding a $T$ with $\tau \ll T \ll n$ $\Delta^2(\bar{x})$ can be estimated as:
\begin{equation}
 \Delta^2(\bar{x})\approx \frac{c_0+2\sum_{k=1}^{T}\left(1-\frac{k}{n}\right) c_k}{n-2 T-1+\frac{T(T+1)}{n}}
\end{equation}
We determine T from the threshold $c_T=0.05\,c_0$. For an uncorrelated time series with $c_k=0$ for $k>0$ and consequently $T=0$, this expression reduces to $\Delta=\sigma/\sqrt{n}$ ($\sigma=\sqrt{c_0}$). 

\section{Distribution functions for selected wavelength intervals}
\label{App:B}
\begin{figure*}
        \centering
        \resizebox{0.95\hsize}{!}{\includegraphics{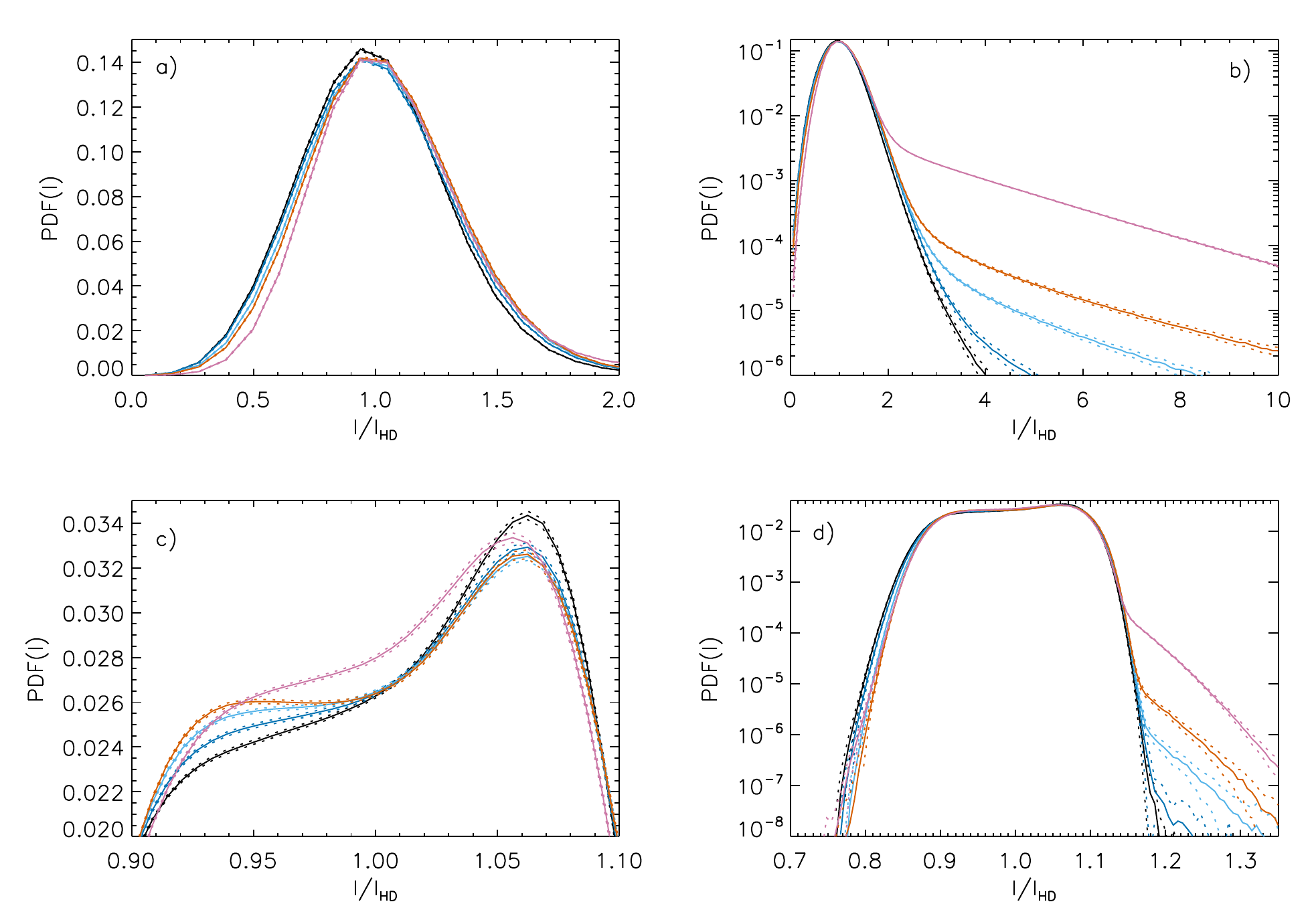}}
        \caption{Probability density functions for the intensity in more narrow wavelength bands for the setups HD (black), SSD$_{\rm WEAK}$ (dark blue), SSD$_{\rm QS}$ (light blue), SSD$_{\rm Strong}$ (orange), and N100 (magenta). Panels a) and b) show the $200 - 300$ nm range and panels c) and d) the $1500 - 2000$ nm range.}
        \label{fig:a1}
\end{figure*}

\begin{figure*}
        \centering
        \resizebox{0.95\hsize}{!}{\includegraphics{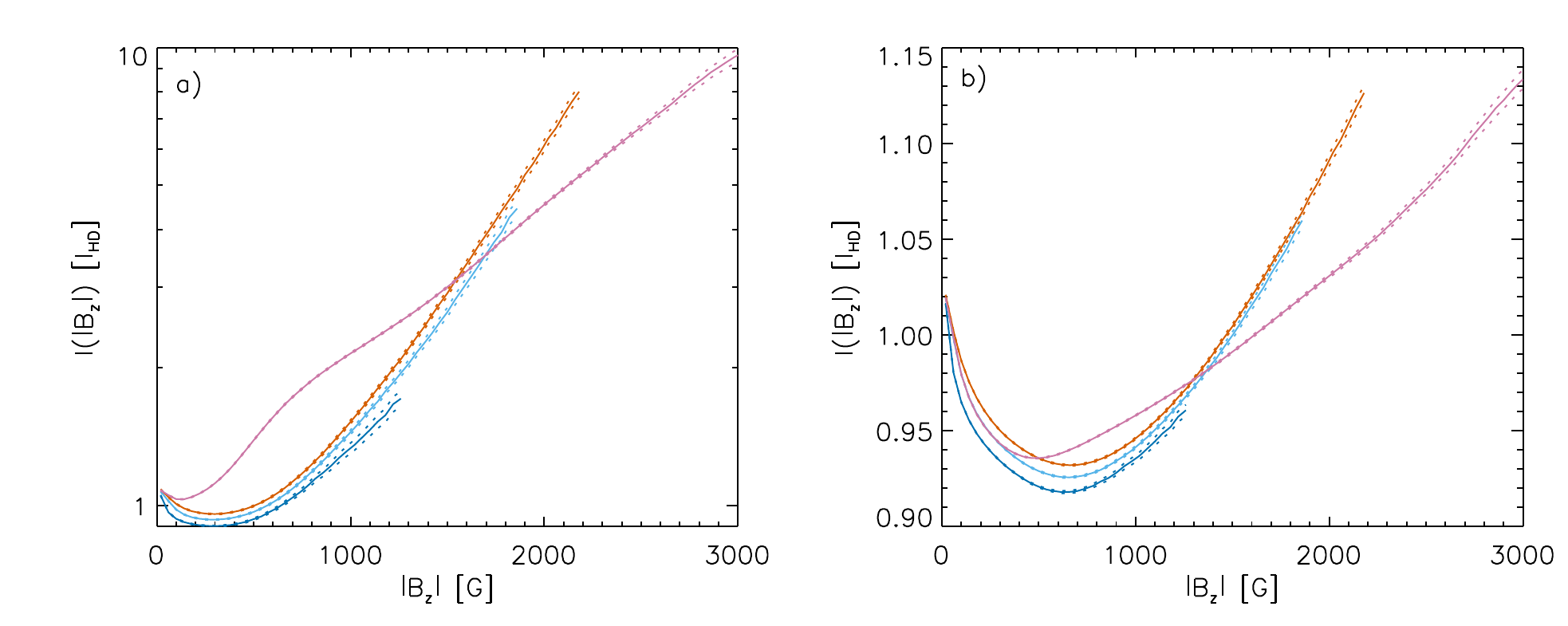}}
        \caption{Relation between vertical field strength and intensity (vertical ray) for the setups HD (black), SSD$_{\rm WEAK}$ (dark blue), SSD$_{\rm QS}$ (light blue), SSD$_{\rm Strong}$ (orange), and N100 (magenta). Panels a) and b) show the $200 - 300$ nm range and panels c) and d) the $1500 - 2000$ nm range.}
        \label{fig:a2}
\end{figure*}

In Section \ref{I_B_histo}, we characterized the radiative properties of the photosphere and their relation to magnetism through probability density functions (PDFs) of bolometric intensity, vertical field strength and their relation. 
For completeness we present here in Figure \ref{fig:a1} the intensity PDFs for the  $200 - 300$ nm and $1500 - 2000$ nm ranges. In the  $200 - 300$ range the core of the intensity PDF is symmetric and the presence of 
magnetic field leads to a systematic shift towards higher intensity. With increasing values of $\umz$ the PDF develops a tail of bright features extending beyond $10\,I_{\rm HD}$ ($I_{\rm HD}$ refers to the reference HD intensity integrated
over the respective wavelength range). In the $1500 - 2000$ nm range the PDF shows 
an asymmetry that is opposite to that found for the bolometric intensity. Furthermore the presence of magnetic field has here the effect of reducing the degree of asymmetry, which is opposite to the behavior found in the bolometric intensity. Figure \ref{fig:a2} presents the relation between pixel-by-pixel intensity and vertical magnetic field strength for the wavelength ranges $200 - 300$ nm and $1500 - 2000$ nm. On a qualitative level these are of similar 
shape to those found for the bolometric intensity in Figure \ref{fig:7}b), the primary difference is found in the intensity values reached for the strongest field (a 10 fold increase in the $200 - 300$ nm range
vs. a 10\% increase for the $1500 - 2000$ nm range). Similar to the bolometric intensity case, pixels in the N100 case are generally brighter (darker)  than the SSD cases for field strength weaker (stronger) than about $1.5$ kG.

\bibliographystyle{aasjournal}
\bibliography{/Users/rempel/Documents/Publications/natbib/papref_m}

\begin{thebibliography}{}
\expandafter\ifx\csname natexlab\endcsname\relax\def\natexlab#1{#1}\fi

\bibitem[{{Buehler} {et~al.}(2013){Buehler}, {Lagg}, \&
  {Solanki}}]{Buehler:2013:cyc_dep}
{Buehler}, D., {Lagg}, A., \& {Solanki}, S.~K. 2013, \aap, 555, A33

\bibitem[{{Castelli} \& {Kurucz}(2004)}]{Castelli:Kurucz:2004}
{Castelli}, F., \& {Kurucz}, R.~L. 2004, arXiv Astrophysics e-prints,
  astro-ph/0405087

\bibitem[{{Criscuoli} {et~al.}(2020){Criscuoli}, {Rempel}, {Haberreiter},
  {Pereira}, {Uitenbroek}, \& {Fabbian}}]{Criscuoli:etal:2020:irrad_codes}
{Criscuoli}, S., {Rempel}, M., {Haberreiter}, M., {et~al.} 2020, {\solphys
  (accepted for publication)}

\bibitem[{{Danilovic} {et~al.}(2016){Danilovic}, {Rempel}, {van Noort}, \&
  {Cameron}}]{Danilovic:2016:SSD_Power}
{Danilovic}, S., {Rempel}, M., {van Noort}, M., \& {Cameron}, R. 2016, \aap,
  594, A103

\bibitem[{{del Pino Alem{\'a}n} {et~al.}(2018){del Pino Alem{\'a}n}, {Trujillo
  Bueno}, {{\v S}t{\v e}p{\'a}n}, \&
  {Shchukina}}]{DelPinoAleman:2018:SSD_Hanle}
{del Pino Alem{\'a}n}, T., {Trujillo Bueno}, J., {{\v S}t{\v e}p{\'a}n}, J., \&
  {Shchukina}, N. 2018, \apj, 863, 164

\bibitem[{{Egorova} {et~al.}(2018){Egorova}, {Schmutz}, {Rozanov}, {Shapiro},
  {Usoskin}, {Beer}, {Tagirov}, \& {Peter}}]{Egorova:etal:2018}
{Egorova}, T., {Schmutz}, W., {Rozanov}, E., {et~al.} 2018, \aap, 615, A85

\bibitem[{{Flyvbjerg} \& {Petersen}(1989)}]{Flyvbjerg:Petersen:1989}
{Flyvbjerg}, H., \& {Petersen}, H.~G. 1989, \jcp, 91, 461

\bibitem[{Gray {et~al.}(2010)Gray, Beer, Geller, Haigh, Lockwood, Matthes,
  Cubasch, Fleitmann, Harrison, Hood, Luterbacher, Meehl, Shindell, van Geel,
  \& White}]{Gray:etal:2010}
Gray, L.~J., Beer, J., Geller, M., {et~al.} 2010, Reviews of Geophysics, 48,
  doi:10.1029/2009RG000282

\bibitem[{{Judge} {et~al.}(2012){Judge}, {Lockwood}, {Radick}, {Henry},
  {Shapiro}, {Schmutz}, \& {Lindsey}}]{Judge:etal:2012}
{Judge}, P.~G., {Lockwood}, G.~W., {Radick}, R.~R., {et~al.} 2012, \aap, 544,
  A88

\bibitem[{{Judge} \& {Saar}(2007)}]{Judge:Saar:2007}
{Judge}, P.~G., \& {Saar}, S.~H. 2007, \apj, 663, 643

\bibitem[{{Khomenko} {et~al.}(2017){Khomenko}, {Vitas}, {Collados}, \& {de
  Vicente}}]{Khomenko:2017:Biermann}
{Khomenko}, E., {Vitas}, N., {Collados}, M., \& {de Vicente}, A. 2017, \aap,
  604, A66

\bibitem[{{Kitiashvili} {et~al.}(2015){Kitiashvili}, {Kosovichev}, {Mansour},
  \& {Wray}}]{Kitiashvili:2015:SSD}
{Kitiashvili}, I.~N., {Kosovichev}, A.~G., {Mansour}, N.~N., \& {Wray}, A.~A.
  2015, \apj, 809, 84

\bibitem[{{Lites}(2011)}]{Lites:2011:hinode_ssd}
{Lites}, B.~W. 2011, \apj, 737, 52

\bibitem[{{Lites} {et~al.}(2014){Lites}, {Centeno}, \&
  {McIntosh}}]{Lites:etal:2014:qs_cyc}
{Lites}, B.~W., {Centeno}, R., \& {McIntosh}, S.~W. 2014, \pasj, 66, S4

\bibitem[{{Lites} {et~al.}(2008){Lites}, {Kubo}, {Socas Navarro}, {Berger},
  {Frank}, {Shine}, {Tarbell}, {Title}, {Ichimoto}, {Katsukawa}, {Tsuneta},
  {Suematsu}, {Shimizu}, \& {Nagata}}]{Lites:etal:2008}
{Lites}, B.~W., {Kubo}, M., {Socas Navarro}, H., {et~al.} 2008, \apj, 672, 1237

\bibitem[{{Meunier}(2018)}]{Meunier:2018:QS_variability}
{Meunier}, N. 2018, \aap, 615, A87

\bibitem[{{Norris} {et~al.}(2017){Norris}, {Beeck}, {Unruh}, {Solanki},
  {Krivova}, \& {Yeo}}]{Norris:etal:2017}
{Norris}, C.~M., {Beeck}, B., {Unruh}, Y.~C., {et~al.} 2017, \aap, 605, A45

\bibitem[{{Peck} {et~al.}(2019){Peck}, {Rast}, {Criscuoli}, \&
  {Rempel}}]{Peck:etal:2019:structure_size}
{Peck}, C.~L., {Rast}, M.~P., {Criscuoli}, S., \& {Rempel}, M. 2019, \apj, 870,
  89

\bibitem[{{Rempel}(2014)}]{Rempel:2014:SSD}
{Rempel}, M. 2014, \apj, 789, 132

\bibitem[{{Rempel}(2018)}]{Rempel:2018:expl}
---. 2018, \apj, 859, 161

\bibitem[{{Schrijver} {et~al.}(2011){Schrijver}, {Livingston}, {Woods}, \&
  {Mewaldt}}]{Schrijver:etal:2011}
{Schrijver}, C.~J., {Livingston}, W.~C., {Woods}, T.~N., \& {Mewaldt}, R.~A.
  2011, \grl, 38, L06701

\bibitem[{{Shapiro} {et~al.}(2011){Shapiro}, {Schmutz}, {Rozanov}, {Schoell},
  {Haberreiter}, {Shapiro}, \& {Nyeki}}]{Shapiro:etal:2011}
{Shapiro}, A.~I., {Schmutz}, W., {Rozanov}, E., {et~al.} 2011, \aap, 529, A67

\bibitem[{{Steinhilber} {et~al.}(2008){Steinhilber}, {Abreu}, \&
  {Beer}}]{Steinhilber:etal:2008}
{Steinhilber}, F., {Abreu}, J.~A., \& {Beer}, J. 2008, Astrophysics and Space
  Sciences Transactions, 4, 1

\bibitem[{{V{\" o}gler} {et~al.}(2005){V{\" o}gler}, {Shelyag}, {Sch{\"
  u}ssler}, {Cattaneo}, {Emonet}, \& {Linde}}]{Voegler:etal:2005}
{V{\" o}gler}, A., {Shelyag}, S., {Sch{\" u}ssler}, M., {et~al.} 2005, \aap,
  429, 335

\bibitem[{{V{\"o}gler} \& {Sch{\"u}ssler}(2007)}]{Voegler:Schuessler:2007}
{V{\"o}gler}, A., \& {Sch{\"u}ssler}, M. 2007, \aap, 465, L43

\bibitem[{{Woods} {et~al.}(2009){Woods}, {Chamberlin}, {Harder}, {Hock},
  {Snow}, {Eparvier}, {Fontenla}, {McClintock}, \&
  {Richard}}]{Woods:etal:2009:WHI}
{Woods}, T.~N., {Chamberlin}, P.~C., {Harder}, J.~W., {et~al.} 2009, \grl, 36,
  L01101

\bibitem[{Wright(2017)}]{pjwright2017}
Wright, P.~J. 2017, ColourBlind: A Collection of IDL Colour-blind-friendly
  Colour Tables,  Zenodo, doi:10.5281/zenodo.840393

\bibitem[{{Wu} {et~al.}(2018){Wu}, {Krivova}, {Solanki}, \&
  {Usoskin}}]{Wu:etal:2018:satire}
{Wu}, C.-J., {Krivova}, N.~A., {Solanki}, S.~K., \& {Usoskin}, I.~G. 2018,
  \aap, 620, A120

\bibitem[{{Yeo} {et~al.}(2017){Yeo}, {Solanki}, {Norris}, {Beeck}, {Unruh}, \&
  {Krivova}}]{Yeo:etal:2017:irrad}
{Yeo}, K.~L., {Solanki}, S.~K., {Norris}, C.~M., {et~al.} 2017, Physical Review
  Letters, 119, arXiv:1709.00920

\bibitem[{{Yeo} {et~al.}(2020){Yeo}, {Solanki}, {Rempel}, {Bhasari}, {Krivova},
  {Shapiro}, {Tagirov}, \& {Witzke}}]{Yeo:etal:2020:Science}
{Yeo}, K.~L., {Solanki}, S.~K., {Rempel}, M., {et~al.} 2020, Science
  (submitted)

\end{thebibliography}

\end{document}